\documentclass[aps,superscriptaddress,onecolumn,11pt]{revtex4-2}

\usepackage{amsmath,amssymb,amsfonts,bbm,graphicx,times}
\usepackage[dvipsnames]{xcolor}
\usepackage{braket}
\usepackage{ntheorem}
\usepackage[normalem]{ulem}
\usepackage[colorlinks=true,bookmarks=false,linkcolor=RoyalBlue,urlcolor=RoyalBlue,citecolor=RoyalBlue,breaklinks]{hyperref}
\usepackage[caption=false, labelfont=bf, labelformat=simple]{subfig}
\usepackage[capitalize]{cleveref}
\usepackage{enumitem}

\newcommand{\ketbra}[2]{\ensuremath{| \: #1 \:\rangle \hspace{-2pt} \langle \: #2 \:  |}}

\newcommand{\eref}[1]{(\ref{#1})}

\newcommand{\llrr}[1]{\ensuremath{\left( #1\right)}}

\DeclareMathOperator{\Tr}{Tr} 

\newtheorem*{theorem-non}{Theorem}
\newtheorem{lemma}{Lemma}
\renewcommand{\phi}{\varphi}
\renewcommand{\epsilon}{\varepsilon}
\renewcommand{\rho}{\varrho}

\usepackage{units}

\begin{document}
\title{Non-perturbative treatment of open-system multi-time expectation values \\ in Gaussian bosonic environments}
\author{Andrea Smirne}
\affiliation{Dipartimento di Fisica ``Aldo Pontremoli'', Universit{\`a} degli Studi di Milano, Via Celoria 16, 20133 Milano-Italy}
\affiliation{Istituto Nazionale di Fisica Nucleare, Sezione di Milano, Via Celoria 16, 20133 Milano-Italy}
\author{Dario Tamascelli}
\affiliation{Dipartimento di Fisica ``Aldo Pontremoli'', Universit{\`a} degli Studi di Milano, Via Celoria 16, 20133 Milano-Italy}
\affiliation{ Institut f{\"u}r Theoretische Physik and Center for Integrated Quantum Science and Technology (IQST), Albert-Einstein-Allee 11, Universit{\"a}t Ulm, 89069 Ulm, Germany}
\author{James Lim} 
\affiliation{ Institut f{\"u}r Theoretische Physik and Center for Integrated Quantum Science and Technology (IQST), Albert-Einstein-Allee 11, Universit{\"a}t Ulm, 89069 Ulm, Germany}
\author{Martin~B. Plenio}
\affiliation{ Institut f{\"u}r Theoretische Physik and Center for Integrated Quantum Science and Technology (IQST), Albert-Einstein-Allee 11, Universit{\"a}t Ulm, 89069 Ulm, Germany}
\author{Susana~F. Huelga} 
\affiliation{ Institut f{\"u}r Theoretische Physik and Center for Integrated Quantum Science and Technology (IQST), Albert-Einstein-Allee 11, Universit{\"a}t Ulm, 89069 Ulm, Germany}

\begin{abstract}
We determine the conditions for the equivalence between the multi-time expectation values of a general finite-dimensional 
open quantum system when interacting with, respectively, an environment undergoing a free unitary evolution
or a discrete environment under a free evolution fixed by a proper Gorini-Kossakowski-Lindblad-Sudarshan generator. 
We prove that the equivalence holds 
if both environments
are bosonic and Gaussian and if the one- and two-time correlation functions of the corresponding interaction operators
are the same at all times. This result leads to a non-perturbative evaluation of the multi-time expectation values of operators and maps of open quantum systems interacting with a continuous
set of bosonic modes by means of a limited number of damped modes,  thus setting the ground for the investigation of open-system multi-time quantities in fully general regimes.
\end{abstract}
\maketitle

\section{Introduction}
The complete statistical description of an open quantum system calls for the characterization of multi-time expectation values  {\cite{breuer02,rivas12}}.  As significant examples, mean values of operators at different times, i.e., multi-time correlation functions, yield optical spectra that are of central
interest in many physical applications \cite{Mukamel1995}, 
while multi-time expectation values of completely positive trace non-increasing maps define the joint
probability distributions associated with sequential measurements at different times \cite{Heinosaari2014}.

The quantum regression theorem \cite{Lax1968,Carmichael1993,gardiner04}
fixes the dynamical equations for multi-time expectation values from those of single-time expectation values, that is, of the quantum state itself.
On the other hand, the validity of the quantum regression theorem  {can be proven only under} quite restrictive conditions \cite{Dumcke1983}, 
which essentially imply that the system-environment correlations created by the
interaction do not impact the multi-time statistics \cite{Swain1981}, or {for} rather specific forms 
of the system-environment coupling \cite{Guarnieri2014,Lonigro2022,Lonigro2022b}. 
Treatments of multi-time expectation values of open quantum systems beyond the quantum regression theorem mainly include the use of perturbative techniques \cite{Alonso2005,Goan2011,Ivanov2015,McCutcheon2016,Ban2018,Holdaway2018,Bonifacio2020}
or the analysis of specific models \cite{Guarnieri2014,Ban2017,Smirne2018,Lonigro2022,Lonigro2022b}, but no general approach able to deal with generic dynamical regimes and physical systems 
is currently available.

\begin{figure}
    \centering
    \includegraphics[width=.6\textwidth]{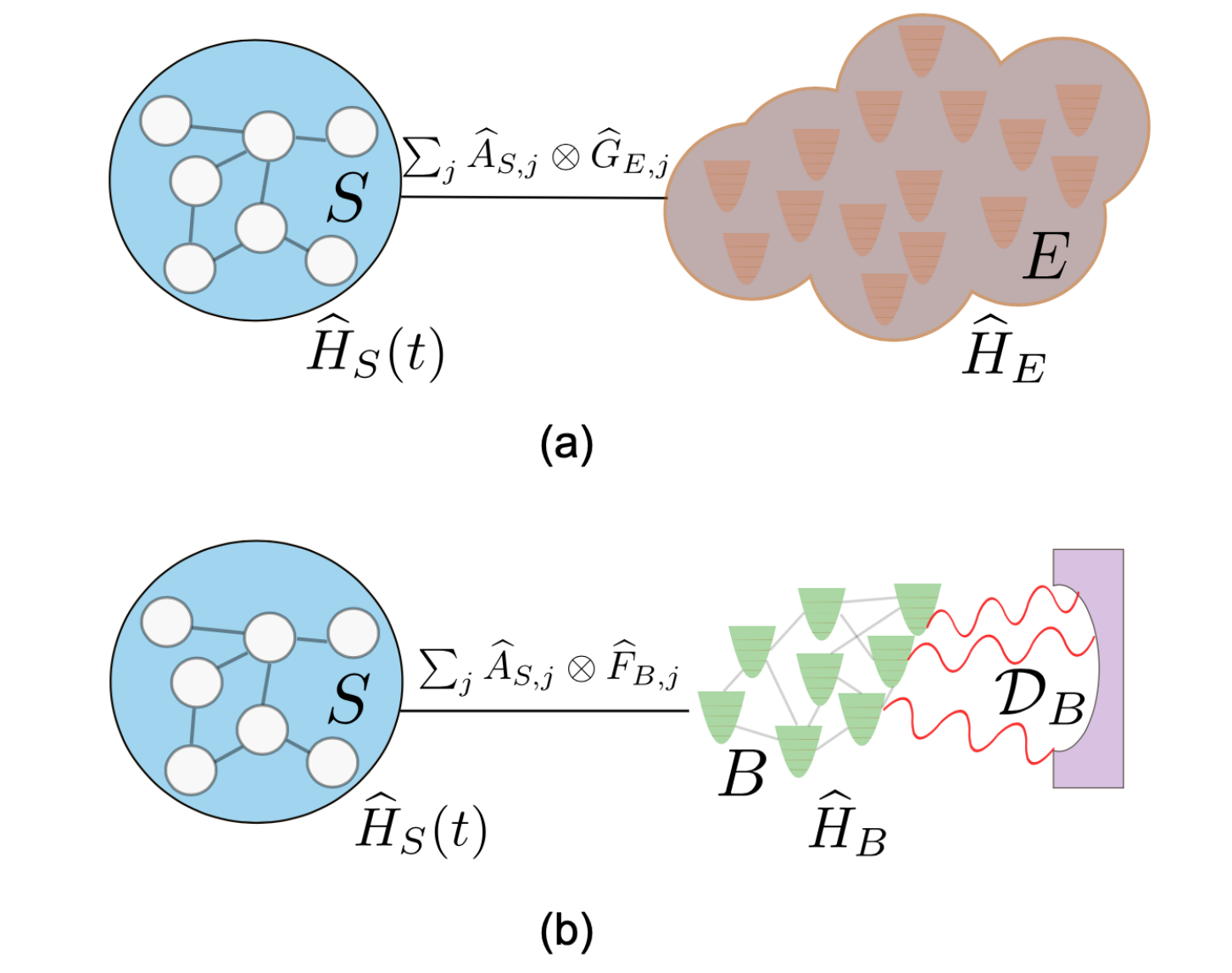}
    \caption{The two configurations we are considering. (a) A system $S$ interacts with a continuum (or countably infinite number) of bosonic modes $E$ undergoing a unitary free evolution. 
    (b) The same system $S$ interacts with a finite number of modes $B$ whose free evolution is determined by a master equation of GKSL type.}
    \label{fig:models}
\end{figure}

Here, we establish a systematic non-perturbative method to evaluate open-system multi-time expectation values by {proving} the equivalence between 
two different system-environment configurations, see Fig.\ref{fig:models}.
In the first one, the system and the (in general, continuous) environment are evolved by a joint unitary dynamics, while in the second one
the same system interacts with an environment consisting of a set of discrete modes and undergoing an evolution
fixed by a Gorini-Kossakowski-Lindblad-Sudarshan (GKLS) generator \cite{Lindblad1976,Gorini1976}. Explicitly, we prove that the multi-time
expectation values
of any sequence of open-system {operators} are identical in the two configurations, provided that the environments are bosonic
and Gaussian, and 
that the one- and two-time correlation functions of the environmental interaction operators coincide under the corresponding free evolutions. 
As a {result, under the conditions set out above, an open quantum system interacting with a continuous set of bosonic modes may be represented equivalently by a discrete set of auxiliary bosonic modes which undergo a GKLS dynamics.}
The specific features of the set of auxiliary damped modes, i.e., their frequencies and couplings, 
can be {determined} via the procedure developed in \cite{mascherpa20}, which relies on a fit of the two-time correlation
functions of the continuous bath with a finite number of exponentials, weighted by possibly complex coefficients. 
Starting from the equivalence theorem proved here, it is thus possible to evaluate in a non-perturbative way multi-time expectation values 
in generic parameter regimes.
This extends the approach put forward in \cite{tamascelli18} and further developed in \cite{mascherpa20}, which concerned the reduced dynamics of the unitary and GKLS configurations and thus ensured the identity between single-time expectation values only; an analogous approach for fermionic baths, addressing both single- and multi-time expectation values has been introduced in \cite{Chen2019}.
Very recently \cite{Nus22}, building upon the equivalence theorem proved here, 2D electronic spectra for a dimeric complex
have been evaluated for a realistic parameter regime.

\section{Unitary and GKLS environments and their multi-time expectation values} \label{sec:lemma}

We start by introducing the two open-system configurations characterized by, respectively, a unitary and a non-unitary (GKLS) free-evolving environment, along with the definitions of the relevant quantities, 
which will be the object of the equivalence theorem presented in the next section.

Consider first any finite dimensional open quantum system $S$
with dimension $d_S$ interacting with
the bosonic environment $E$ according to the global Hamiltonian
\begin{equation}\label{eq:hse}
\widehat{H}_{SE}(t) = \widehat{H}_S(t)+\widehat{H}_E  + \sum^{\kappa}_{j=1} \widehat{A}_{S, j} \otimes \widehat{G}_{E,j},
\end{equation}
where  {$\kappa\leq d_S^2 $,} and we allow for a possible time dependence in the free system-Hamiltonian
$\widehat{H}_S(t)$, so that  {general time dependent controls} are included in our treatment; $\widehat{H}_E$ is instead the Hamiltonian of the bosonic environment on which we do not make, for the time being, any particular assumptions. Moreover, we assume
an initial  factorized state,
\begin{equation}\label{eq:rse0}
\rho_{SE}(0) = \rho_S(0) \otimes \rho_E(0),
\end{equation}
so that the dynamics on $S$ is completely positive \cite{breuer02}.
Our main focus is the general (time-ordered) $n$-time expectation value
\begin{eqnarray}
&&\langle O_n(t_n) \ldots O_1(t_1); O'_1(t_1)\ldots O'_n(t_n)\rangle_U =  \label{eq:oou0} \\
&&\hspace{2.cm}\Tr_{SE}\left\{\mathcal{U}^{\dagger}_{SE}(t_n)\left[\widehat{O}_n\right]
\ldots \mathcal{U}^{\dagger}_{SE}(t_1)\left[\widehat{O}_1\right]\rho_{SE}(0)
\mathcal{U}^{\dagger}_{SE}(t_1)\left[\widehat{O}'_1 \right]\ldots
\mathcal{U}^{\dagger}_{SE}(t_n)\left[\widehat{O}'_n \right]
\right\},\notag
\end{eqnarray}
with $t_n\geq \ldots \geq t_1\geq 0$, where $\widehat{O}_k$ is a short-hand notation for  {$\widehat{O}_k \otimes \mathbbm{1}_E $,}
$\widehat{O}_1, \ldots  {,} \widehat{O}_n$ and $ \widehat{O}'_1 {,} \ldots {,} \widehat{O}'_n$ are generic
open-system operators and $\mathcal{U}^{\dagger}_{SE}(t)$ defines the $S-E$ unitary evolution
in the Heisenberg picture
\begin{equation}\label{eq:c11}
    \mathcal{U}^{\dagger}_{SE}(t)[\widehat{O}] =\widehat{U}^{\dagger}_{SE}(t)
    \widehat{O} \widehat{U}_{SE}(t), \qquad \widehat{U}_{SE}(t)=T_{\leftarrow} e^{-i\int_0^t d s\widehat{H}_{SE}(s)},
\end{equation} 
with $T_{\leftarrow}$ the chronological time-ordering operator.
Note that Eq.(\ref{eq:oou0}) defines the most general multi-time
expectation values appearing in the description of the evolution of open quantum systems \cite{Accardi1982,Dumcke1983,Li2018};
for $\widehat{O}'_1=\ldots=\widehat{O}'_n=\mathbbm{1}_S$
and self-adjoint $\widehat{O}_1 \ldots \widehat{O}_n$ we recover the multi-time
correlation functions of the observables associated with the latter operators \cite{breuer02},
while for $\widehat{O}'_1 = \widehat{O}^\dag_1 {,} \ldots  {,} \widehat{O}'_n =\widehat{O}^\dag_n$
we obtain the multi-time probabilities associated with a generic sequence of measurements \cite{Heinosaari2014,Milz2020};
finally, identity operators appearing alternatively among the left and the right operators
lead us to the definition of signals in multi-dimensional spectroscopy \cite{Nus22}.
Introducing the unitary propagators in the Schr{\"o}dinger picture 
\begin{equation}
    \mathcal{U}_{SE}(t,t')[\rho_{SE}]=  \widehat{U}_{SE}(t,t') \rho_{SE} \widehat{U}^{\dagger}_{SE}(t,t'), \qquad
    \widehat{U}_{SE}(t,t') = \widehat{U}_{SE}(t)\widehat{U}^\dagger_{SE}(t')= T_{\leftarrow} e^{-i\int_{t'}^t d s\widehat{H}_{SE}(s)},
\end{equation}
the definition in Eq.(\ref{eq:oou0})
can be expressed equivalently as
\begin{eqnarray}
&&\langle O_n(t_n) \ldots O_1(t_1); O'_1(t_1)\ldots O'_n(t_n)\rangle_U =  \label{eq:oou} \\
&&\hspace{2.cm}\Tr_{SE}\left\{\widehat{O}_n\mathcal{U}_{SE}(t_n,t_{n-1})\left[\ldots
\mathcal{U}_{SE}(t_2,t_{1})\left[\widehat{O}_1\mathcal{U}_{SE}(t_1)\left[ \rho_{SE}(0) \right]
\widehat{O}'_1\right]\ldots
\right]
\widehat{O}'_n
\right\}.\notag
\end{eqnarray}
In addition, we consider the single-time expectation values of the environmental interaction terms,
\begin{equation}
G^U_{E,j}(t)=\Tr_E \left\{e^{i \widehat{H}_{E} t}\widehat{G}_{E,j} e^{-i \widehat{H}_{E} t}\rho_{E}(0)\right\},
\end{equation}
as well as the two-time correlation functions
\begin{eqnarray}
C^U_{j j'}(t+s,s) & = \Tr_E\left \{e^{i \widehat{H}_{E} (t+s)}\widehat{G}_{E,j}e^{-i \widehat{H}_{E} (t+s)} 
e^{i \widehat{H}_{E} s} \widehat{G}_{E,j'} e^{-i \widehat{H}_{E} s} \rho_E(0)\right\}. \label{eq:corru}
\end{eqnarray}
Crucially, the definitions of
$G^U_{E,j}(t)$ and $C^U_{j j'}(t+s,s)$
correspond to the general definitions in Eq.(\ref{eq:oou0}) but with respect to
the free evolution of the environment, as fixed by $\widehat{H}_{E}$; thus, their explicit
evaluation does not involve the solution of the full dynamics.\\

The second configuration we take into account consists of the same quantum system $S$ interacting with a bosonic environment $B$, such that the global evolution is fixed by the GKLS generator 
\begin{eqnarray}
    \mathcal{L}_{SB}(t)[\rho_{SB}] &=& 
  -i \left[\widehat{H}_{SB}(t), \rho_{SB} \right]   + \mathcal{D}_B[\rho_{SB} ] = \nonumber \\
    && -i \left[\widehat{H}_{SB}(t), \rho_{SB} \right]  \label{eq:lin} 
    +\sum_{j=1}^\ell \gamma_j \llrr{ \widehat{L}_{B,j} \rho_{SB} \widehat{L}_{B,j}^\dagger
    -\frac{1}{2} \left \{ \widehat{L}_{B,j}^\dagger \widehat{L}_{B,j}, \rho_{SB} \right \}}, 
\end{eqnarray}
where the Hamiltonian term can be decomposed as 
\begin{align}
\widehat{H}_{SB}(t) = \widehat{H}_S(t) +  \widehat{H}_B + \sum^{\kappa}_{j=1}
\widehat{A}_{S, j} \otimes \widehat{F}_{B,j},
\label{eq:hamL}
\end{align}
with $\widehat{H}_B$ the (yet unspecified) free-environment Hamiltonian, 
and $\gamma_j \geq 0$, ensuring the complete
positivity of the evolution on $S+B$ \cite{Lindblad1976,Gorini1976,breuer02}; 
note that $\ell$ is related with the number of degrees of freedom of B, and hence
with the complexity of the non-unitary configuration. 
We stress that the dissipative term acts 
on $B$ only. Moreover, the free Hamiltonian of the open system  {${\hat H}_S(t)$},
as well as the operators $\widehat{A}_{S,j}$ acting on the open system side in the interaction part of the Hamiltonian $\widehat{H}_{SB}(t)$,
are the same as those in $\widehat{H}_{SE}(t)$ (compare \cref{eq:hse} and \cref{eq:hamL}): in this sense, the open system is the same in the two configurations. 
We further assume a factorized initial state
\begin{equation}\label{eq:rsb0}
\rho_{SB}(0) = \rho_S(0) \otimes \rho_B(0),
\end{equation}
which guarantees the complete positivity of the evolution on $S$.
The general $n$-time expectation values
under the GKLS evolution are defined as
\begin{widetext}
\begin{eqnarray}
&&\langle O_n(t_n) \ldots O_1(t_1); O'_1(t_1)\ldots O'_n(t_n)\rangle_L =  \label{eq:ooL}\\
&&
\hspace{2.cm}\Tr_{SB}\left\{\widehat{O}_n \Lambda_{SB}(t_n,t_{n-1})\left[\ldots 
\Lambda_{SB}(t_2,t_1)\left[\widehat{O}_1\Lambda_{SB}(t_1)\left[\rho_{SB}(0)\right]
\widehat{O}'_1\right]\ldots
\right]
\widehat{O}'_n \right\},\notag
\end{eqnarray}
\end{widetext}
with the propagators
\begin{equation}
\Lambda_{SB}(t,t') = T_{\leftarrow}e^{\int_{t'}^t\mathcal{L}_{SB}(s)}
\end{equation}
with $\Lambda_{SB}(t)=\Lambda_{SB}(t,0)$;
$\widehat{O}_k$ is now a short-hand notation for $\widehat{O}_k \otimes \mathbbm{1}_B$.
As before, we also consider the one- and two-time correlation functions of the environmental interaction operators
with respect to the ``free'' evolution
of the environment (i.e., without taking into account the presence of the system $S$),
which is now fixed by the GKLS generator on $B$
\begin{equation}\label{eq:ldag}
\mathcal{L}_B [\rho_B] = - i \left[\widehat{H}_B, \rho_B \right] + \mathcal{D}_B[\rho_B ],
\end{equation}
so that we have
\begin{equation}
F^L_{B,j}(t)=\Tr_B \left\{\widehat{F}_{B,j} e^{\mathcal{L}_B t}\left[\rho_{B}(0)\right]\right\},
\end{equation}
as well as the two-time correlation functions
\begin{align}
    C_{j j'}^L(t+s,s) = \Tr_B \left\{\widehat{F}_{B,j}e^{\mathcal{L}_{B}t}\left[ \widehat{F}_{B,j'} e^{\mathcal{L}_B s}\left[
\rho_{B}(0)\right]\right]\right\}.
\label{eq:cljj}
\end{align}

As stated earlier, we will restrict our attention to the case where both the environments $E$ and $B$ are 
Gaussian. In particular, we assume that
\begin{enumerate}[label=\roman*)]
    \item   both the initial environmental states $\rho_E(0)$
and $\rho_B(0)$ are Gaussian states,
\item the environmental free Hamiltonians $\widehat{H}_E$ and $\widehat{H}_B$ 
are at most quadratic
in the corresponding creation and annihilation operators,
\item the environmental coupling operators $\widehat{G}_{E,j}$ and $\widehat{F}_{B,j}$, as well as
the Lindblad operators $\widehat{L}_{B,j}$ are linear in the creation and annihilation operators.
\end{enumerate}
Crucially, these assumptions imply
that the $n$-time correlation functions of the environmental interaction operators under the free evolution
of the environment (generalizing
the definitions in, respectively, Eq.(\ref{eq:corru}) and Eq.(\ref{eq:cljj})) can be expressed in terms of the 
corresponding one- and two-time correlation functions; 
in other terms, the environmental interaction operators
are Gaussian operators with respect to the free dynamics \cite{Efremov1981}. 
In fact, assumptions ii) and iii) guarantee that the environmental interaction operators evolved via the free environmental evolution at a generic time
are linear in the creation and annihilation operators,
so that, using also assumption i), one can apply the Wick's theorem and then express any environmental $n$-time correlation function
in terms of one- and two-time correlation functions~\cite{bramberger20}.

\section{Equivalence theorem}

We can now formulate the equivalence theorem, which equates the open-system multi-time expectation values
of the unitary configuration and those of
the GKLS configuration, provided that the one- and two-time
environmental correlation functions coincide.
\begin{theorem-non}
Consider the same open system $S$ in the two configurations characterized by, respectively, Eqs.~(\ref{eq:hse})-(\ref{eq:corru}) and Eqs.~(\ref{eq:lin})-(\ref{eq:cljj}), and
with the same initial state $\rho_S(0)$. If the bosonic environments $E$ and $B$ satisfy the assumptions i)-iii) above,
and if 
\begin{equation}\label{eq:ass1}
    F^L_{B,j}(t)=  G^U_{E,j}(t) \,\,\qquad \forall j \, \mbox{and } t\geq 0
\end{equation}
and
\begin{equation}\label{eq:ass2}
    C^L_{j j'}(t+s,s) = C^U_{j j'}(t+s,s)  \,\,\qquad \forall j,j' \mbox{ and } \forall t, s\geq 0,
\end{equation}
then  
\begin{eqnarray}
    &&\langle O_n(t_n) \ldots O_1(t_1); O'_1(t_1)\ldots O'_n(t_n)\rangle_L 
    = \langle O_n(t_n) \ldots O_1(t_1); O'_1(t_1)\ldots O'_n(t_n)\rangle_U 
    \notag\\
    && 
    \,\,\qquad \forall O_1, \ldots O_n, ; O'_1\ldots O'_n 
    \,\mbox{ and }\, \forall t_n\geq \ldots \geq t_1\geq 0.\label{eq:thether}
\end{eqnarray}
\end{theorem-non}
\textbf{Proof.}
The proof essentially follows the same steps of the proof for the one-time expectation values derived in \cite{tamascelli18},
generalizing it to the multi-time case by means of induction, analogously to what is done in \cite{Chen2019} for the fermionic case.

The key point is the introduction of a third configuration, which represents the unitary dilation \cite{gardiner04} of the non-unitary configuration.
Hence, consider the tripartite system $S-B-\tilde{E}$ given by the $S-B$ system introduced above and a continuous distribution
of bosonic modes labelled by $\tilde{E}$. The tripartite system evolves unitarily by means of the Hamiltonian \cite{foot3}
\begin{align}
\widehat{H}_{SB\tilde{E}}(t) &= \widehat{H}_{SB}(t)+\widehat{H}_{\tilde{E}}  + \widehat{V}_{B\tilde{E}}, \nonumber\\
\widehat{H}_{\tilde{E}} &=  \sum^{\ell}_{j=1} \int_{-\infty}^{\infty} d \omega\,  \omega \,  
\widehat{b}^{\dag}_{\tilde{E}}(\omega, j) \widehat{b}_{\tilde{E}}(\omega, j), \label{eq:hx}\\
\widehat{V}_{B\tilde{E}} &= \sum^{\ell}_{j=1} \sqrt{-\frac{\gamma_j}{2 \pi}}
\int_{-\infty}^{\infty}\hspace{-5pt} d \omega 
\widehat{L}_{B,j} \widehat{b}^{\dag}_{\tilde{E}}(\omega, j) - \widehat{L}^{\dag}_{B,j}
\widehat{b}_{\tilde{E}}(\omega, j), \nonumber 
\end{align}
where $\widehat{b}_{\tilde{E}}(\omega, j)$ and $\widehat{b}^{\dag}_{\tilde{E}}(\omega, j)$ are bosonic
annihilation and creation operators of a Fock space $\mathcal{H}_{\tilde{E},j}$,
\begin{equation}\label{eq:CCR}
\left[\widehat{b}_{\tilde{E}}(\omega, j), \widehat{b}^{\dag}_{\tilde{E}}(\omega', j')\right] = \delta_{j j'} \delta(\omega-\omega');
\end{equation}
the initial state is
\begin{equation}\label{eq:vava}
\rho_{SB\tilde{E}}(0) = \rho_{SB}(0) \otimes \ketbra{0_{\tilde{E}}}{0_{\tilde{E}}},
\end{equation}
with $\rho_{SB}(0)$ as in \cref{eq:rsb0} and $\ket{0_{\tilde{E}}} =  \otimes_{j=1}^\ell  \ket{0_j}$ the vacuum state of
the global Fock space associated with $\tilde{E}$, $\mathcal{H}_{\tilde{E}} = \otimes_{j=1}^\ell \mathcal{H}_{\tilde{E},j}$. 
The $n$-time expectation values of such a unitary evolution can be defined as (compare with Eq.(\ref{eq:oou}))
\begin{eqnarray}\label{eq:oox}
&&\langle O_n(t_n) \ldots O_1(t_1); O'_1(t_1)\ldots O'_n(t_n)\rangle_X =   \\
&&\hspace{2.cm}\Tr_{SB\tilde{E}}\left\{\widehat{O}_n\widetilde{\mathcal{U}}(t_n,t_{n-1})\left[
\ldots
\widetilde{\mathcal{U}}(t_2,t_{1})\left[\widehat{O}_1\widetilde{\mathcal{U}}(t_1)\left[ \rho_{SB\tilde{E}}(0) \right]
\widehat{O}'_1\right]\ldots\right]
\widehat{O}'_n
\right\},\notag
\end{eqnarray}
where $\widehat{O}_k$ is now a short-hand notation for $\widehat{O}_k\otimes\mathbbm{1}_{B\tilde{E}}$, and the subscript $X$ indicates that the expectation is taken over the extended $SB\tilde{E}$ space. Here
$\widetilde{\mathcal{U}}(t,t')$ is the propagator of the unitary dynamics on $S-B-\tilde{E}$ which reads
\begin{equation}\label{eq:propsbet}
    \widetilde{\mathcal{U}}(t,t')\left[\rho_{SB\tilde{E}}\right] = 
\widehat{U}_{SB\tilde{E}}(t,t') \rho_{SB\tilde{E}} \widehat{U}^{\dagger}_{SB\tilde{E}}(t,t'), \qquad
    \widehat{U}_{SB\tilde{E}}(t,t') = T_{\leftarrow} e^{-i\int_{t'}^t d s\widehat{H}_{SB\tilde{E}}(s)};
\end{equation}
finally, the one- and two-time correlation functions of the environmental interaction operators (that are still
the $\widehat{F}_{B,j}$, as the interaction with $S$ is still given by $\widehat{H}_{SB}(t)$)
with respect to the free unitary dynamics on $B-\tilde{E}$ fixed by the Hamiltonian 
\begin{equation}\label{eq:accab}
    \widehat{H}_{B\tilde{E}} = \widehat{H}_{B}+\widehat{H}_{\tilde{E}} +\widehat{V}_{B\tilde{E}}
    \end{equation}
are    
\begin{equation}
    F^X_{B,j}(t) = \Tr_{B\tilde{E}} \left\{e^{i \widehat{H}_{B\tilde{E}} t}\widehat{F}_{B,j}\otimes \mathbbm{1}_{\tilde{E}} e^{-i \widehat{H}_{B\tilde{E}} t}\rho_{B\tilde{E}}(0)\right\}
\end{equation}
and
\begin{align}
C^X_{j j'}(t+s,s) &= \Tr_{B\tilde{E}}\left\{e^{i \widehat{H}_{B\tilde{E}} (t+s)}\widehat{F}_{B,j}\otimes \mathbbm{1}_{\tilde{E}}e^{-i
    \widehat{H}_{B\tilde{E}}(t+s)} 
    e^{i \widehat{H}_{B\tilde{E}}s}  \widehat{F}_{B,j'}\otimes \mathbbm{1}_{\tilde{E}}e^{-i \widehat{H}_{B\tilde{E}}s} \rho_{B\tilde{E}}(0)\right\}.\nonumber 
\end{align}

To prove the theorem, the first observation is that the multi-time expectation values 
of two unitary configurations with the same open system $S$ and initial state $\rho_S(0)$ coincide if the corresponding $n$-time correlation functions of the environmental
interaction operators under the respective free dynamics are the same;
this can be shown, for example, via the Dyson expansion \cite{tamascelli18}.
But then, since we are dealing with Gaussian environments \cite{foot}
-- so that the $n$-time correlation functions are uniquely determined by the one- and two-time correlation functions -- we immediately have the following implication
\begin{align}
\left.
  \begin{tabular}{ccc}
$F^X_{B,j}(t)$& $=$ &$G^U_{E,j}(t)$ \\
$C^X_{j j'}(t+s,s)$  & $=$ & $C^U_{j j'}(t+s,s)$
  \end{tabular}
\right\}&
\,\,\, \forall j, j' \, \forall t,s\geq 0 \nonumber \\
&\hspace{-5cm} \Longrightarrow \quad 
 \langle O_n(t_n) \ldots O_1(t_1); O'_1(t_1)\ldots O'_n(t_n)\rangle_X 
    = \langle O_n(t_n) \ldots O_1(t_1); O'_1(t_1)\ldots O'_n(t_n)\rangle_U 
    \notag\\
    & 
    \,\,\qquad \forall O_1, \ldots O_n, ; O'_1\ldots O'_n 
    \,\mbox{ and }\, \forall t_n\geq \ldots \geq t_1\geq 0.
 \label{eq:impl}
\end{align}

Now, we can directly use Lemma 1 and Lemma 2 of \cite{tamascelli18}, which trace back to the validity
of the quantum regression theorem
for the reduced dynamics on $B$ obtained via the partial trace over the unitary dynamics on $B-\tilde{E}$. As to ease the reading, we report these two lemmas, adapted to the current notation. 

\textbf{Lemma 1.} Consider a tripartite system $S-B-\tilde{E}$ undergoing the unitary evolution fixed by Eq.~\eref{eq:hx}
with initial state $\rho_{SR\tilde{E}}(0)$ set as in Eq.~\eref{eq:vava}. 
Denoting as $\rho_{SB}^X(t)$ the reduced $S-B$ state at time $t$, i.e.,
\begin{equation}\label{eq:rhox}
\rho_{SB}^X(t) = \Tr_{\tilde{E}}\left\{e^{-i \hat{H}_{SB\tilde{E}}  t}\left(\rho_{SB}(0) \otimes
\ketbra{0_{\tilde{E}}}{0_{\tilde{E}}}\right) e^{i \hat{H}_{SB\tilde{E}}  t}\right\},
\end{equation}
then (recalling that $\rho_{SB}(t)$ is the state whose evolution is fixed by the GKLS generator defined in Eq.\eqref{eq:lin})
\begin{equation}\label{eq:step1}
\rho_{SB}^X(t) = \rho_{SB}(t).
\end{equation}
\textbf{Lemma 2.} 
Given the unitary dynamics on $B-\tilde{E}$ fixed by the Hamiltonian of Eq.~\eref{eq:accab}, its correlation functions satisfy
\begin{equation}\label{eq:step2}
C^X_{j j'}(t+s,s) =C^L_{j j'}(t+s,s) \quad \forall t,s \geq 0.
\end{equation}

In fact, Lemma 2 tells us that the two-time correlation functions
of operators on $B$ with respect to the unitary dynamics on $B-\tilde{E}$
coincide with those obtained by means of the GKLS dynamics on $B$,
i.e., $C^X_{j j'}(t+s,s)=C^L_{j j'}(t+s,s)$, while the analogous
correspondence between the expectation values, $F^X_{B,j}(t)=F^L_{B,j}(t) $, is implied directly by Lemma 1 \cite{foot2}.
Replacing these identities in \cref{eq:impl}, we thus have
\begin{align}
\left.
  \begin{tabular}{ccc}
$F^L_{B,j}(t)$& $=$ &$G^U_{E,j}(t)$ \\
$C^L_{j j'}(t+s,s)$  & $=$ & $C^U_{j j'}(t+s,s)$
  \end{tabular}
\right\}&
\,\,\, \forall j, j' \, \forall t,s\geq 0 \nonumber \\
&\hspace{-5cm} \Longrightarrow \quad 
 \langle O_n(t_n) \ldots O_1(t_1); O'_1(t_1)\ldots O'_n(t_n)\rangle_X 
    = \langle O_n(t_n) \ldots O_1(t_1); O'_1(t_1)\ldots O'_n(t_n)\rangle_U 
    \notag\\
    & 
    \,\,\qquad \forall O_1, \ldots O_n, ; O'_1\ldots O'_n 
    \,\mbox{ and }\, \forall t_n\geq \ldots \geq t_1\geq 0, 
\end{align}
so that to prove the theorem, we only need to show the
identity 
\begin{align}
  &   \langle O_n(t_n) \ldots O_1(t_1); O'_1(t_1)\ldots O'_n(t_n)\rangle_X 
    = \langle O_n(t_n) \ldots O_1(t_1); O'_1(t_1)\ldots O'_n(t_n)\rangle_L 
    \notag\\
    & 
    \,\,\qquad \forall O_1, \ldots O_n, ; O'_1\ldots O'_n 
    \,\mbox{ and }\, \forall t_n\geq \ldots \geq t_1\geq 0, \label{eq:atlast}
\end{align}    
which directly follows from the next Lemma.

\begin{lemma}\label{lem:lem}
Given the two configurations characterized by, respectively, Eqs.~(\ref{eq:lin}), (\ref{eq:hamL}), (\ref{eq:ooL})-(\ref{eq:ldag})
and Eqs.~(\ref{eq:hx})-(\ref{eq:accab}), with the same $S-B$ system and generic initial state $\rho_{SB}(0)$,
for any set of operators 
$\widehat{D}_1, \ldots \widehat{D}_n, \widehat{D}'_1, \ldots \widehat{D}'_n$ on $\mathcal{H}_{SB}$, the Hilbert space associated with the system $S-B$, and 
times $t_n \geq \ldots \geq t_1 \geq 0$
we have 
\begin{align}
  &   \langle D_n(t_n) \ldots D_1(t_1); D'_1(t_1)\ldots D'_n(t_n)\rangle_X 
    = \langle D_n(t_n) \ldots D_1(t_1); D'_1(t_1)\ldots D'_n(t_n)\rangle_L. \label{eq:fin}\end{align}  
\end{lemma}

Indeed, \cref{eq:atlast} follows from this Lemma by setting $\rho_{SB}(0)$ as in \cref{eq:rsb0} and
$\widehat{D}_1 = \widehat{O}_1 \otimes \mathbbm{1}_B, \ldots \widehat{D}'_n = \widehat{O}'_n \otimes \mathbbm{1}_B$,
(compare \cref{eq:fin} with Eqs.(\ref{eq:ooL}) and (\ref{eq:oox})), which concludes the proof of the Theorem.\\

\noindent
\textbf{Proof of Lemma \ref{lem:lem}}. Essentially, we generalize the proof of Lemma 1 in \cite{tamascelli18} by means of induction.

For $n=1$, \cref{eq:fin} is equivalent to Lemma 1.
Let us then assume that \cref{eq:fin} holds for $n=k-1$ and show that this implies its validity for $n=k$.
Actually, the validity of \cref{eq:fin} for $n=k$ is equivalent to the validity of
the differential equation, recalling the
definitions in Eqs.(\ref{eq:ooL}) and (\ref{eq:oox})
and introducing
$\tau=t_k-t_{k-1}$,
\begin{align}
 & \frac{d}{d\tau} \Tr_{SB\tilde{E}}\left\{\widehat{D}_k\widetilde{\mathcal{U}}(\tau+t_{k-1},t_{k-1})\left[\widehat{D}_{k-1}
\widetilde{\mathcal{U}}(t_{k-1},t_{k-2})\left[\ldots
\widehat{D}_{1}\widetilde{\mathcal{U}}(t_1)\left[ \rho_{SB\tilde{E}}(0) \right]
\widehat{D}'_{1}\ldots\right]\widehat{D}'_{k-1}
\right]\widehat{D}'_{k}
\right\} \notag\\
& =\frac{d}{d\tau}\Tr_{SB}\left\{\widehat{D}_k \Lambda_{SB}(\tau+t_{k-1},t_{k-1})\left[\widehat{D}_{k-1}
\Lambda_{SB}(t_{k-1},t_{k-2})\left[\ldots
\widehat{D}_{1}\Lambda_{SB}(t_1)\left[ \rho_{SB}(0) \right]
\widehat{D}'_{1}\ldots\right]\widehat{D}'_{k-1}
\right]\widehat{D}'_{k}
\right\}
    \label{eq:exex}
\end{align}
for any $\widehat{D}_1, \ldots, \widehat{D}_{k},\widehat{D}'_1, \ldots, \widehat{D}'_{k} $
and $t_{k-1}\geq \ldots \geq t_{1}\geq 0, \tau \geq 0$
along with the initial conditions for $\tau=0$
(and for any $t_{k-1}\geq\ldots \geq t_1\geq 0$ and operators 
$\widehat{D}_1, \ldots, \widehat{D}_{k},\widehat{D}'_1, \ldots, \widehat{D}'_{k}$)
\begin{align}
 & \Tr_{SB\tilde{E}}\left\{\widehat{D}_k \widehat{D}_{k-1}
\widetilde{\mathcal{U}}(t_{k-1},t_{k-2})\left[\ldots
\widehat{D}_{1}\widetilde{\mathcal{U}}(t_1)\left[ \rho_{SB\tilde{E}}(0) \right]
\widehat{D}'_{1}\ldots\right]\widehat{D}'_{k-1}\widehat{D}'_{k}
\right\} \notag\\
& =\Tr_{SB}\left\{\widehat{D}_k \widehat{D}_{k-1}
\Lambda_{SB}(t_{k-1},t_{k-2})\left[\ldots
\widehat{D}_{1}\Lambda_{SB}(t_1)\left[ \rho_{SB}(0) \right]
\widehat{D}'_{1}\ldots\right]\widehat{D}'_{k-1} \widehat{D}'_{k}
\right\};
    \label{eq:exexic}
\end{align}
in both the previous equations, we implied the tensor product of the operators $\widehat{D}$ and $\widehat{D}'$ with the identities,
namely with $\mathbbm{1}_{B\tilde{E}}$ at the left hand side and $\mathbbm{1}_{B}$ at 
the right hand side.
The initial conditions hold by the induction hypothesis
for the $S-B$ operators $\widehat{D}_1, \ldots {,} \widehat{D}_{k-1} \widehat{D}_k$,
$\widehat{D}'_1,  \ldots, \widehat{D}'_{k-1} \widehat{D}'_k$, and thus we only have to prove the validity of \cref{eq:exex}.

Introducing the $S-B-\tilde{E}$ unitary evolution operator in the Heisenberg picture (compare with Eq.(\ref{eq:propsbet}))
\begin{equation}
    \widetilde{\mathcal{U}}^\dagger(t)\left[\rho_{SB\tilde{E}}\right] = \widehat{U}_{SB\tilde{E}}^\dagger \rho_{SB\tilde{E}} \widehat{U}_{SB\tilde{E}} \qquad \widehat{U}_{SB\tilde{E}}(t) = T_{\leftarrow}e^{-i \int_0^t ds \widehat{H}_{SB\tilde{E}}}(s),\label{eq:utild}
\end{equation}
we can rewrite the left hand side of \cref{eq:exex} as (compare with Eq.(\ref{eq:oou0}))
\begin{align}
 &  \frac{d}{d\tau} \Tr_{SB\tilde{E}}\left\{\widehat{D}_k\widetilde{\mathcal{U}}(\tau+t_{k-1},t_{k-1})\left[\widehat{D}_{k-1}
\widetilde{\mathcal{U}}(t_{k-1},t_{k-2})\left[\ldots
\widehat{D}_{1}\widetilde{\mathcal{U}}(t_1)\left[ \rho_{SB\tilde{E}}(0) \right]
\widehat{D}'_{1}\ldots\right]\widehat{D}'_{k-1}
\right]\widehat{D}'_{k}
\right\}\notag\\
 = & \frac{d}{d\tau} \Tr_{SB\tilde{E}}\left\{
 \widetilde{\mathcal{U}}^\dag(\tau+t_{k-1})\left[
 \widehat{D}'_k \widehat{D}_k \right]
 \widetilde{\mathcal{U}}^\dag(t_{k-1})\left[
\widehat{D}_{k-1} \right] 
 \ldots
  \widetilde{\mathcal{U}}^\dag(t_1)\left[\widehat{D}_1\right]
  \rho_{SB\tilde{E}}(0)
  \widetilde{\mathcal{U}}^\dag(t_1)\left[\widehat{D}'_1\right]
  \ldots
   \widetilde{\mathcal{U}}^\dag(t_{k-1})\left[
\widehat{D}'_{k-1} \right]
\right\}. \notag
\end{align}
The Hamiltonian defined in \cref{eq:hx} implies that the derivative
with respect to $\tau$ takes the form \cite{gardiner04}
\begin{align}
    & \frac{d}{d\tau} \Tr_{SB\tilde{E}}\left\{
 \widetilde{\mathcal{U}}^\dag(\tau+t_{k-1})\left[
 \widehat{D}'_k \widehat{D}_k \right]
 \widetilde{\mathcal{U}}^\dag(t_{k-1})\left[
\widehat{D}_{k-1} \right] 
 \ldots
  \widetilde{\mathcal{U}}^\dag(t_1)\left[\widehat{D}_1\right]
  \rho_{SB\tilde{E}}(0)
  \widetilde{\mathcal{U}}^\dag(t_1)\left[\widehat{D}'_1\right]
  \ldots
   \widetilde{\mathcal{U}}^\dag(t_{k-1})\left[
\widehat{D}'_{k-1} \right]
\right\} \notag\\
&= Tr_{SB\tilde{E}}\Bigg\{ \Bigg( i  \left[\widehat{\widetilde{H}}_{SB}(\tau+t_{k-1}), \widehat{\widetilde{D}}'_{k}(\tau+t_{k-1})\widehat{\widetilde{D}}_{k}(\tau+t_{k-1})\right]
\notag\\ &
-\sum_{j=1}^{\ell}\left[\widehat{\widetilde{D}}'_{k}(\tau+t_{k-1})\widehat{\widetilde{D}}_{k}(\tau+t_{k-1}),\widehat{\widetilde{L}}^{\dag}_{B,j}(\tau+t_{k-1})\right]
\left(\frac{\gamma_j}{2} \widehat{\widetilde{L}}_{B,j}(\tau+t_{k-1})+\sqrt{\gamma_j} \widehat{b}_{in}(\tau+t_{k-1}, j) \right) \nonumber\\
&+\sum_{j=1}^{\ell}\left(\frac{\gamma_j}{2} \widehat{\widetilde{L}}^{\dag}_{B,j}(\tau+t_{k-1})+\sqrt{\gamma_j}
\widehat{b}^{\dag}_{in}(\tau+t_{k-1}, j) \right) \left[\widehat{\widetilde{D}}'_{k}(\tau+t_{k-1})\widehat{\widetilde{D}}_{k}(\tau+t_{k-1}),\widehat{\widetilde{L}}_{B,j}(\tau+t_{k-1})\right]\Bigg)
\notag\\& \times 
\widehat{\widetilde{D}}_{k-1}(t_{k-1})\ldots \widehat{\widetilde{D}}_1(t_1)\rho_{SB\tilde{E}}(0)\widehat{\widetilde{D}}'_1(t_1)\ldots 
\widehat{\widetilde{D}}'_{k-1}(t_{k-1}) \Bigg\},\label{eq:mobb}
\end{align}
where
\begin{equation}\label{eq:mava}
\widehat{\widetilde{O}}(t) = \widetilde{\mathcal{U}}^\dag(t)\big[\widehat{O}(t) \big],
\end{equation}
for every possibly time-dependent operator $\widehat{O}(t)$ on $\mathcal{H}_{SB\tilde{E}}$ in the Schr{\"o}dinger picture;
in addition, crucially, we introduced the input field \cite{gardiner04}
\begin{equation}
\widehat{b}_{in}(t, j) = \frac{1}{\sqrt{2 \pi}} \int_{-\infty}^{\infty} d \omega e^{- i \omega t} \widehat{b}_{\tilde{E}}(\omega, j)
\end{equation}
that is defined on $\mathcal{H}_{\tilde{E}}$, satisfies the commutation relations
\begin{align}\label{eq:inou}
&\left[\widehat{b}_{in}(t,j), \widehat{b}^{\dag}_{in}(s,j')\right] = \delta_{j j'}\delta(t-s), \notag\\ 
&\left[\widehat{b}_{in}(t,j), \widehat{b}_{in}(s,j')\right] = 0 \quad \forall t,s \geq 0
\end{align}
and acts as an annihilation operator on the vacuum,
\begin{equation}\label{eq:zer}
\widehat{b}_{in}(t, j) \ket{0_{\tilde{E}}} = 0 .
\end{equation}
Moreover, since any $\widehat{\widetilde{D}}(t)$ depends on $b_{in}(s,j)$ only for times $s\leq t$ (as can be seen from the Heisenberg equation of
$\widehat{\widetilde{D}}(t)$ \cite{tamascelli18}), \cref{eq:inou} implies
\begin{equation}\label{eq:aqq}
    \left[\widehat{b}_{in}(t+s,j), \widehat{\widetilde{D}}(t)(s) \right] = 0 \quad \forall j, s, t>0.
\end{equation}
Thus, at the right-hand side of \cref{eq:mobb}, for any $\tau>0$, we can exchange the input field $\widehat{b}_{in}(\tau+t_{k-1},j)$
and the product of operators $\widehat{\widetilde{D}}_{k-1}(t_{k-1})\ldots \widehat{\widetilde{D}}_1(t_1)$, so that because of 
\cref{eq:vava,eq:zer} the term with $\widehat{b}_{in}(\tau+t_{k-1},j)$ is equal to zero. Analogously,
using the cyclicity of the trace
we can see that the term with $\widehat{b}^\dag_{in}(t+t_{k-1},j)$ vanishes as well, since
the latter commutes with the product $\widehat{\widetilde{D}}'_{1}(t_{1})\ldots \widehat{\widetilde{D}}'_{k-1}(t_{k-1})$
and $\bra{0}_{\tilde{E}}\widehat{b}^\dagger_{in}(t,j)=0$.
All in all, using once again \cref{eq:mava} we are left with
\begin{align}
    &  \frac{d}{d\tau} \Tr_{SB\tilde{E}}\left\{\widehat{D}_k\widetilde{\mathcal{U}}(\tau+t_{k-1},t_{k-1})\left[\widehat{D}_{k-1}
\widetilde{\mathcal{U}}(t_{k-1},t_{k-2})\left[\ldots
\widehat{D}_{1}\widetilde{\mathcal{U}}(t_1)\left[ \rho_{SB\tilde{E}}(0) \right]
\widehat{D}'_{1}\ldots\right]\widehat{D}'_{k-1}
\right]\widehat{D}'_{k}
\right\} \notag\\
&= Tr_{SB\tilde{E}}\Big\{ 
\widetilde{\mathcal{U}}^\dag(\tau+t_{k-1})\left[
\mathcal{L}^\dag_{SB}(\tau+t_{k-1})\left[\widehat{D}'_k\widehat{D}_k\right]\right]
\widehat{\widetilde{D}}_{k-1}(t_{k-1})\ldots \widehat{\widetilde{D}}_1(t_1)\rho_{SB\tilde{E}}(0)
\widehat{\widetilde{D}}'_1(t_1) \ldots \widehat{\widetilde{D}}'_{k-1}(t_{k-1}) \Big\},\label{eq:ququ}
\end{align}
where we introduced the dual of the GKLS generator \cite{breuer02} in Eq.(\ref{eq:lin}):
   \begin{eqnarray}
    \mathcal{L}^\dag_{SB}(t)[\widehat{D}] &=& i \left[\widehat{H}_{SB}(t), \widehat{D} \right]  \label{eq:lindual}
    +\sum_{j=1}^\ell \gamma_j \llrr{ \widehat{L}^\dagger_{B,j} \widehat{D} \widehat{L}_{B,j}
    -\frac{1}{2} \left \{ \widehat{L}_{B,j}^\dagger \widehat{L}_{B,j}, \widehat{D} \right \}}; \notag
\end{eqnarray}
note that the derivative of the expectation value in \cref{eq:ququ} is defined by continuity at $\tau=0$.

We can now proceed analogously to the final part of the proof of Lemma 2 in \cite{tamascelli18}.
Consider the basis of operators on $\mathcal{H}_{SB}$ given by $\left\{\widehat{M}_{a b, SB} = \ketbra{\varphi_a}{\varphi_b}\right\}_{a,b}$, 
where $\left\{\ket{\varphi_a}\right\}_{a}$ is a basis of the Hilbert space $\mathcal{H}_{SB}$,
so that one has \cite{Carmichael1993}
\begin{equation}\label{eq:lmat}
\mathcal{L}^{\dag}_{SB}(t)\left[\widehat{M}_{a b, SB} \right] = \sum_{c d}  W_{ab}^{cd}(t) \widehat{M}_{c d, SB}
\end{equation}
with
\begin{equation}
    W_{ab}^{cd}(t) = \bra{\varphi_c}\mathcal{L}^{\dag}_{SB}(t)\left[\ketbra{\varphi_{a}}{\varphi_b}\right] \ket{\varphi_d}.
\end{equation}
Looking at the corresponding $n$-time expectation values, 
\cref{eq:ququ,eq:lmat} for $\widehat{D}'_k\widehat{D}_k=\widehat{M}_{a b,SB}$ imply
\begin{eqnarray}
&&\frac{d}{d\tau} \Tr_{SB\tilde{E}}\left\{\widehat{M}_{a b, SB} \widetilde{\mathcal{U}}(\tau+t_{k-1},t_{k-1})\left[\widehat{D}_{k-1}
\widetilde{\mathcal{U}}(t_{k-1},t_{k-2})\left[\ldots
\widehat{D}_{1}\widetilde{\mathcal{U}}(t_1)\left[ \rho_{SB\tilde{E}}(0) \right]
\widehat{D}'_{1}\ldots\right]\widehat{D}'_{k-1}
\right]
\right\} \notag\\
&&= \sum_{c d} W_{ab}^{cd}(\tau+t_{k+1})\label{eq:aa1}\\
&&\times\Tr_{SB\tilde{E}}\left\{\widehat{M}_{c d, SB} \widetilde{\mathcal{U}}(\tau+t_{k-1},t_{k-1})\left[\widehat{D}_{k-1}
\widetilde{\mathcal{U}}(t_{k-1},t_{k-2})\left[\ldots
\widehat{D}_{1}\widetilde{\mathcal{U}}(t_1)\left[ \rho_{SB\tilde{E}}(0) \right]
\widehat{D}'_{1}\ldots\right]\widehat{D}'_{k-1}
\right]
\right\}. \notag
\end{eqnarray}
Using the dual in \cref{eq:lmat}, one can readily show that the right-hand side of \cref{eq:exex},
evaluated for $\widehat{D}'_k\widehat{D}_k=\widehat{M}_{a b,SB}$, is
\begin{align}
& \frac{d}{d\tau}\Tr_{SB}\left\{\widehat{M}_{a b, SB} \Lambda_{SB}(\tau+t_{k-1},t_{k-1})\left[\widehat{D}_{k-1}
\Lambda_{SB}(t_{k-1},t_{k-2})\left[\ldots
\widehat{D}_{1}\Lambda_{SB}(t_1)\left[ \rho_{SB}(0) \right]
\widehat{D}'_{1}\ldots\right]\widehat{D}'_{k-1}
\right]
\right\}  \notag\\
& = \sum_{c d}  W_{ab}^{cd}(\tau+t_{k-1}) \label{eq:aa2}\\
&\times \Tr_{SB}\left\{\widehat{M}_{c d, SB}  \Lambda_{SB}(\tau+t_{k-1},t_{k-1})\left[\widehat{D}_{k-1} \Lambda_{SB}(t_{k-1},t_{k-2}) 
\left[
\ldots
\widehat{D}_1 
\Lambda_{SB}(t_1) \rho_{SB}(0) \right]\right]
\right\}. \notag
\end{align}
Comparing Eq.(\ref{eq:aa1}) and Eq.(\ref{eq:aa2}), we see that the $k$-time expectation values
involving the operator-basis elements $\left\{\widehat{M}_{a b, SB}\right\}_{a,b}$ (and all the other operators 
$\widehat{D}_1, \ldots, \widehat{D}_{k-1},\widehat{D}'_1, \ldots, \widehat{D}'_{k-1}$ and times are fixed)
satisfy the same system of differential equations in the two configurations, implying the validity of \cref{eq:exex}
for a generic product $\widehat{D}'_k\widehat{D}_k$, and hence for 
any couple of operators
$\widehat{D}_k$ and $\widehat{D}'_k$ (and any $\widehat{D}_1, \ldots, \widehat{D}_{k-1},\widehat{D}'_1, \ldots, \widehat{D}'_{k-1}$, $t_{k-1}\geq \ldots \geq t_1\geq 0$ and $t\geq 0$);
this concludes our proof.

\section{Discussion and conclusions}
We have proven that general multi-time expectation values of an open quantum system interacting with a unitary Gaussian bosonic environment are  {identical} to those of the same open quantum system interacting with a GKLS Gaussian bosonic environment,
provided that the one- and two-time correlation functions of the bath interaction operators under the respective free dynamics
are the same. 
While such an equivalence between two unitary Gaussian environments directly follows from the very notion of Gaussiantiy, which allows one to express all the bath multi-time correlation functions in terms of one- and two-time correlation functions, when comparing a unitary and a GKLS environment two further steps are needed. Firstly, one introduces a unitary
dilation, which involves a continuous set of bosonic modes with a flat (i.e., not frequency-dependent) distribution of the couplings, see Eqs.(\ref{eq:hx}) and (\ref{eq:CCR}),
and which yields exactly the GKLS dynamics when the continuous bosonic modes are traced out, see Lemma 1.
Secondly, one shows the equivalence between the multi-time expectation values of the configuration with a GKLS bath and those of the corresponding dilation, see Lemma \ref{lem:lem}. Let us stress that, in fact, this means that the unitary dilation guarantees the exact validity of the quantum regression theorem for the multi-time expectation values of the GKLS configuration,
that is, they can be equivalently evaluated with the unitary dilated dynamics or with the reduced dynamical maps that are obtained
after tracing out the continuous flat bosonic bath, see in particular Eq.(\ref{eq:exex}).

For any finite-dimensional open quantum system, the equivalence theorem proved here, along with the one in \cite{Chen2019}, concludes
the proof of the equivalence 
between a unitary and a GKLS environment, when they are Gaussian and either bosonic or fermionic.
These results thus set the ground for a non-perturbative evaluation of multi-time quantities, when combined with the strategy
put forward in \cite{mascherpa20} to derive explicitly the parameters fixing an auxiliary configuration consisting of a small
number of damped modes.
In the case of structured baths, with both fermionic and bosonic components, the equivalence can still be guaranteed, as long
as the two configurations are structured in the same way with respect to the corresponding fermionic and bosonic partitions.
The investigation of infinite dimensional open quantum systems looks indeed much more demanding,
and a natural starting point could be provided by the evolutions preserving the Gaussianity of the global state 
\cite{Grabert1988,Hu1992}.
Other two scenarios that go beyond the range of applicability of the equivalence theorems proved here
are represented by the presence of initial system-environment correlations and by out-of-time-order correlators (OTOCs).
In the former case, it is generally not possible to describe the evolution of the open system via a single (time-dependent) reduced dynamical map defined on the whole set of open-system states \cite{Pechukas1994,Alicki1995,Stelmachovic2001,Jordan2006,Brodutch2013,Buscemi2014,Vacchini2014,Paz2019,Trevisan2021}, and thus the possible extension of the equivalence theorems to the case with initial correlations would likely call for a definition of positivity domains or multiple maps
that is consistent in both the unitary- and GKLS-bath configurations. OTOCs represent multi-time expectation values where
times are not ordered, and it is a priori not-clear that a GKLS configuration can be properly defined in this case;
however, we note that a proper extension of the quantum regression theorem can be indeed formulated for OTOCs \cite{Blocher2019}.

As last remark, we point out that, as mentioned in \cite{tamascelli18}, the approach that led us to prove the equivalence theorem might be useful even in the presence of non-Gaussian baths \cite{bramberger20,Sung2019}. Here a general, not model-dependent, equivalence between the unitary and GKLS configurations would call for the equality of the $n$-time correlation functions of the two baths for an arbitrary $n$, thus being unrealistic to verify in practical situations. On the other hand, relying on perturbative techniques in which the $n$-time correlation functions appear only starting from
the $n$-th order of the expansion \cite{breuer02}, by controlling the equality of the first $n$-time correlation functions one can ensure that the unitary and GKLS configurations give the same predictions up to the truncation error of the perturbative expansion at order $n+1$, which can be estimated with standard techniques~\cite{breuer02}.  

\acknowledgments 
It is our pleasure to dedicate this manuscript to the memory of Professor Kossakowski. His work has been a constant inspiration for us. We feel honored to have known him personally and will never forget his kindness and his class. \\

AS and DT authors acknowledge
support from UniMi, via PSR-2 2020 and PSR-2 2021.  {MBP and SFH acknowledge support by the DFG via QuantERA project ExtraQt.}

\bibliography{biblio.bib}

\begin{thebibliography}{48}
\bibitem [{foo({\natexlab{a}})}]{foot3}%
  \BibitemOpen
  \href@noop {} {} ({\natexlab{a}}),\ \bibinfo {note} {{The} Hamiltonian
  $\widehat{H}_{SB\tilde{E}}(t)$ is indeed singular, due to the flat
  distribution of the couplings; as usual within the input-output formalism, a
  proper unitary evolution associated with it can be formulated via the quantum
  stochastic calculus \cite{gardiner04}.}\BibitemShut {Stop}%
\bibitem [{foo({\natexlab{b}})}]{foot}%
  \BibitemOpen
  \href@noop {} {} ({\natexlab{b}}),\ \bibinfo {note} {{The} original unitary
  system $S-E$ is Gaussian by hypothesis, while the unitary dilation
  $S-B-\tilde{E}$ is Gaussian due to the hypotheses on $S-B$; in fact, i)
  $\rho_{B\tilde{E}}(0) = \rho_B(0)\otimes
  \ketbra{0_{\tilde{E}}}{0_{\tilde{E}}}$ is Gaussian since both $\rho_B(0)$ and
  the vacuum are so; ii) the free environmental Hamiltonian
  $\widehat{H}_{B\tilde{E}}$ in Eq.(\ref{eq:accab}) is quadratic since
  $\widehat{H}_B$ and $\widehat{H}_{\tilde{E}}$ are so, and since the Lindblad
  operators $\widehat{L}_{B,j}$ are linear in the creation and annihilation
  operators (see the definition of $\widehat{V}_{B\tilde{E}}$ in
  Eq.(\ref{eq:hx})); iii) the environmental interaction operators
  $\widehat{F}_{B,j}$ are linear in the creation and annihilation
  operators.}\BibitemShut {Stop}%
\bibitem [{foo({\natexlab{c}})}]{foot2}%
  \BibitemOpen
  \href@noop {} {} ({\natexlab{c}}),\ \bibinfo {note} {{Lemma} 1 has been
  proved in \cite{tamascelli18} for a time-independent Hamiltonian; however, it
  is easy to see that the same proof holds for a time-dependent
  $\widehat{H}_S(t)$ by using the Heisenberg-picture Hamiltonian as defined by
  Eq.(\ref{eq:utild}); on the other hand, Lemma 2 only involves the systems $B$
  and $B-\tilde{E}$ and it is thus not affected at all by the time-dependence
  on $S$.}\BibitemShut {Stop}%
\end{thebibliography}

\end{document}